\documentclass[reprint, amsmath, amssymb, aps, superscriptaddress]{revtex4-1}

\usepackage{graphicx, dcolumn,subfigure,bm}

\usepackage{hyperref}

\begin{document}

\title{A self-calibrating polarimeter to measure Stokes parameters}

\author{V. Andreev}
\affiliation{Department of Physics, Harvard University, Cambridge, Massachusetts 02138, USA}
\affiliation{Technische Universit\"at M\"unchen, Physik-Department, D-85748 Garching, Germany}
\author{C. D. Panda}
\affiliation{Department of Physics, Harvard University, Cambridge, Massachusetts 02138, USA}
\author{P. W. Hess}
\affiliation{Department of Physics, Harvard University, Cambridge, Massachusetts 02138, USA}
\author{B. Spaun}
\affiliation{Department of Physics, Harvard University, Cambridge, Massachusetts 02138, USA}
\author{G. Gabrielse}
\affiliation{Department of Physics, Harvard University, Cambridge, Massachusetts 02138, USA}

%\date{February 28, 2017}
\date{February 19, 2021}

\begin{abstract}

An easily constructed and operated polarimeter precisely determines the relative Stokes parameters that characterize the polarization of laser light. The polarimeter is calibrated in situ without removing or realigning its optical elements, and it is largely immune to fluctuations in the laser beam intensity.  The polarimeter's usefulness is illustrated by measuring thermally-induced birefringence in the indium-tin-oxide coated glass field plates used to produce a static electric field in the ACME collaboration's measurement of the electron electric dipole moment.

\end{abstract}

\maketitle

\section{Introduction}

Light polarimetry remains extremely important in many fields of physics \cite{OverviewPol}.  Measurements of the polarization of light reveal information about interactions between excited states of atoms \cite{AtomicPhysicsPol}.  Among many applications in astronomy \cite{AstronomyPolar}, the polarization of light from interstellar dust reveals the magnetic field that aligns the dust \cite{DustAlignment,DustMagnField}. 
Light polarization also probes the magnetic field in the plasmas used for nuclear fusion studies, insofar as magnetic fields cause the Faraday rotation of linear light polarization and Cotton-Mouton changes in light ellipticity \cite{ITERPolarimeter}.  For the most precise measurement of the electron's electric dipole moment, polarimetry of the thermally-induced circular polarization gradient within glass electric field plates was crucial for understanding the mechanisms that dominantly contributed to the systematic uncertainty \cite{Baron:2013eja}.

Drawing upon the early work of G. G. Stokes \cite{Stokes, stokes1852composition}, and a later experimental realization \cite{Berry:77}, we investigate the limits of a rotating waveplate polarimeter for determining the polarization state of partially polarized laser light.  The design is easy to realize and is robust in its operation.  With a calibration procedure introduced here, it is straightforward to internally calibrate the polarimeter without the need to remove or realign optical elements. 
The polarimeter is designed to be largely immune to fluctuations in light intensity, and it has been used at intensities up to a $100~\rm{mW}/\rm{mm}^2$. The relative fractions of circularly polarized and linearly polarized light can typically be measured with uncertainties below $0.1\,\%$ and $0.4\,\%$, respectively.  

Light polarization can be measured in various ways \cite{Goldstein}.  
Polarimeters similar to ours, but lacking the internal calibration mechanism and immunity to  intensity fluctuations, can handle up to several $\rm{mW}/\rm{mm}^2$ \cite{HindsPol, ThorlabsPol} and attain uncertainties less than $\pm 0.9\%$ in the Stokes parameters;  they have even been recommended for student labs \cite{AJP}.  Lower precision is also typically attained using other measurement methods.  Light is sometimes split to travel along optical paths with differing optical elements, the polarization state being deduced from the relative intensities transmitted along the paths  \cite{Azzam:92, Azzam:85, Peinado:13, He:15}. Alternatively, the light can be analyzed using optical elements whose properties vary spatially,  with the polarization revealed by the spatially varying intensity \cite{Chang:14, Zhao:14, Lepetit}.  

The usefulness of our internally calibrated polarimeter is demonstrated by characterizing a circular polarization gradient across a nominally linearly polarized laser beam. This gradient is produced by thermally-induced birefringence caused by the high intensity of the laser light traveling through glass electric field plates coated with an electrically conducting layer of indium tin oxide.  Such spatial polarization gradient contributes substantially to the systematic uncertainty in the first-generation ACME measurement of the electron's electric dipole moment \cite{Baron:2013eja}. The small and well-characterized uncertainties of the polarimeter make it possible to characterize new glass electric field plates that were designed to produce much smaller spatial polarization gradients in the second-generation ACME apparatus.

 This paper is structured as follows: After reviewing the Stokes parameters in Section \ref{sec:stokes} and introducing the basics of a rotating waveplate polarimeter in Section \ref{sec:polarimeter}, we describe its laboratory realization together with the intensity normalization scheme in Section \ref{sec:realization}. Section \ref{sec:extractingStokes} summarizes how to extract the Stokes parameters from a polarimeter measurement. The calibration technique we developed and our analysis of the uncertainties is presented in Section \ref{sec:calibuncert}. Finally, we illustrate the performance of the polarimeter with an ellipticity gradient measurement in Section \ref{sec:application}.
 
\section{Stokes parameters}
\label{sec:stokes}
At any instant point in space and time, the electric field of a light wave points in a particular direction. If the electric field follows a repeatable path during its oscillations, the light wave is said to be polarized. Averaged over some time that is long compared to the oscillation period of the light, however, the light may be only partially polarized or even completely unpolarized if the direction of the electric field varies in a non-periodic way.  
Stokes showed that fully polarized light and partially polarized light can be characterized, in principle, by intensities transmitted after the light passes through each of four simple configurations of optical elements:  

\small
\begin{subequations}
	\label{eq:originalStokes}
	\begin{align}
		I =& I(0^\circ) + I(90^\circ) =  I(45^\circ) + I(-45^\circ)  \nonumber  \\
		   =&  I_\text{RHC} + I_\text{LHC}, \\
		M =& I(0^\circ) - I(90^\circ),  \\
		C =& I(45^\circ) - I(-45^\circ), \\
		S =& I_\text{RHC} - I_\text{LHC}.
	\end{align}
\end{subequations}
\normalsize
The total intensity $I$, and the two linear polarizations $M$ and $C$, are given in terms of intensities $I(\alpha)$ measured after the light passes through a perfect linear polarizer whose transmission axis is oriented at an angle $\alpha$ with respect to the polarization of the incoming light. The circular polarization $S$ is the difference between the intensity of right- and left-handed circularly polarized light, $I_\text{RHC}$ and $I_\text{LHC}$, that, as we shall see, can be deduced using a quarter-waveplate followed by a linear polarizer \cite{Goldstein}.

\subsection{Fully polarized light}
Elliptical polarization is the most general state of a fully polarized plane wave traveling in the $z$ direction with frequency $\omega$ and wavenumber $k$.  In cartesian coordinates, the electric field is 
\begin{equation}
\label{eq:EfieldPlaneWave}
   \bm{\vec{ \mathcal{E}}} =  \hat{\bf x}\,  \mathcal{E}_{0x}\cos(\omega t - kz+\phi)  + \hat{\bf y} \,\mathcal{E}_{0y}\cos(\omega t - kz ), 
\end{equation}
where $ \mathcal{E}_{0x}$ and $ \mathcal{E}_{0y}$ are the absolute values of orthogonal electric field components. $\phi$ represents the phase difference between the two orthogonal components. The Stokes vector, defined with respect to the polarization measured in the plane perpendicular to the propagation direction $\hat{k}$, is
\begin{equation}
\vec{S}=\begin{pmatrix} I \\ M \\ C \\ S \end{pmatrix} =\begin{pmatrix}    \mathcal{E}_{0x}^2 +  \mathcal{E}_{0y}^2  \\  \mathcal{E}_{0x}^2  -   \mathcal{E}_{0y}^2 \\ 2 \,  \mathcal{E}_{0x}  \mathcal{E}_{0y} \cos \phi   \\  2 \,  \mathcal{E}_{0x}  \mathcal{E}_{0y} \sin \phi    \end{pmatrix} %2 \, \text{Re} \langle E_x E_y^* \rangle   \\ - 2 \text{Im}  \langle E_x E_y^*\rangle  \end{pmatrix} 
\label{eq:Stokes}
\end{equation} 
whereupon it follows that
\small
\begin{equation}
I^2 = M^2+C^2+S^2
\label{eq:imcs}
\end{equation}
\normalsize
relates the four Stokes parameters in this case.  

While $I$ describes the total light intensity, the dimensionless quantities $M/I$, $C/I$ and $S/I$ determine the polarization state of light.  The linear polarization fraction is  $L/I= \sqrt{(M/I)^2+(C/I)^2}$ and the circular polarization fraction is $S/I$, with 
\begin{equation}
(L/I)^2+(S/I)^2 =1.
\label{eq:LiSi}
\end{equation}
Because the relative intensities are summed in quadrature, nearly complete linear polarization  (e.g.\ $L/I = 99 \%$) corresponds to a circular polarization that is still substantial (e.g.\ $S/I = 14 \%$). 

\begin{figure}[tb!]
\includegraphics[width=0.7\hsize,keepaspectratio]{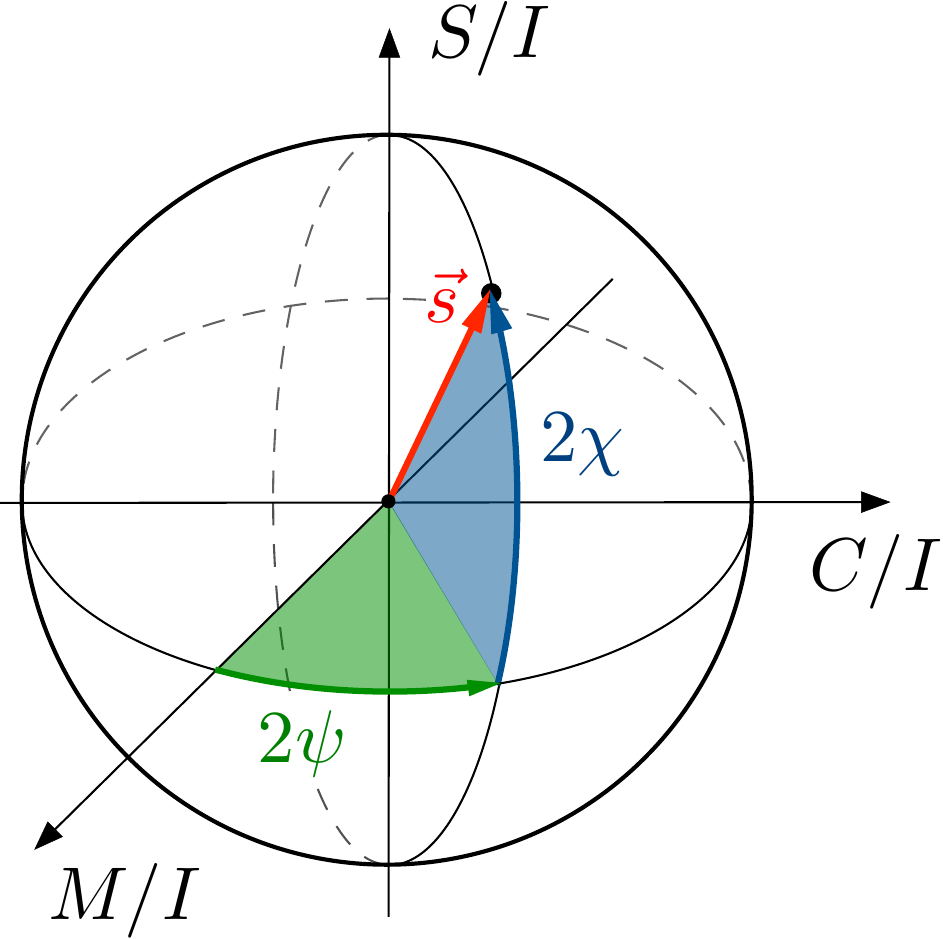}
\caption{The three relative Stokes parameters on three orthogonal axes trace out the Poincar\'{e} sphere, with each point on the surface a possible state of fully polarized light.}  
\label{fig:poincare}
\end{figure}

Points on the Poincar\'{e} sphere (Fig. \ref{fig:poincare}) represent the elliptical polarization state with a relative Stokes vector  
\begin{equation}
\label{eq:vectorSi}
\vec{s}=\begin{pmatrix} M/I \\ C/I \\ S/I \end{pmatrix} =  \begin{pmatrix} \cos 2 \chi \cos 2 \psi  \\  \cos 2 \chi \sin 2 \psi   \\ \sin 2 \chi  \end{pmatrix}.
\end{equation}
The linear rotation angle is defined by $\tan 2 \psi = C/M$ and the ellipticity angle is defined by $S/I=\sin 2 \chi $.

\subsection{Partially Polarized Light}

The light is partially polarized if the amplitudes $\mathcal{E}_{0x}$ and $\mathcal{E}_{0y}$ and the phase 
$\phi$ fluctuate enough so that an average over time reduces the size of the average correlations between electric field components. The Stokes parameters are then defined by the time averages of Eq.~\ref{eq:Stokes}, with the averaging interval being long compared to both the oscillation period and the inverse bandwidth of the Fourier components that describe the light.  The unpolarized part of the light contributes only to the first of the four Stokes parameters, $I$, and not to $M$, $C$ or $S$. The polarization fraction that survives the averaging, $P$, is given by
\begin{equation}
 P^2 =(M/I)^2+(C/I)^2+(S/I)^2 \le 1.
\end{equation}
When $P=1$, the light is completely polarized and the polarization vector describes a point on the Poincar\'{e} sphere. For partially polarized light, the length of the polarization vector is shortened such that it will now describe a point inside the sphere. When $P=0$, the light is fully unpolarized.

\section{Rotating Waveplate Polarimeter}
\label{sec:polarimeter}

\begin{figure}[tb!]
	\includegraphics[width=\hsize,keepaspectratio]{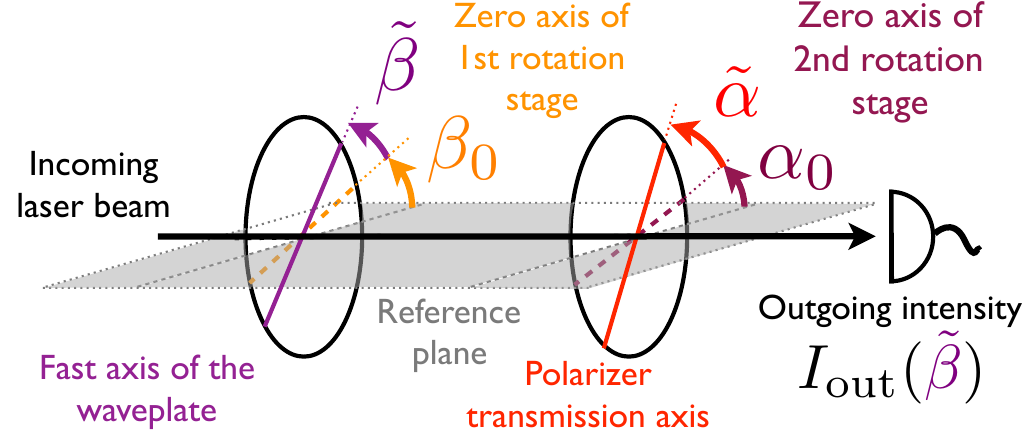}
	\caption{A rotating waveplate polarimeter comprised of a rotatable waveplate followed by a linear polarizer and a detector.  The axes of the optical elements are specified with respect to a reference plane.  }
	\label{fig:genscheme}
\end{figure}

The general scheme of a rotating waveplate polarimeter is shown in Fig. \ref{fig:genscheme}.  Light travels first through a quarter-waveplate which can be rotated to determine the polarization state. The light then travels through a linear polarizer which can be rotated to internally calibrate the angular location of the fast axis of the waveplate and the orientation angle of the linear polarizer transmission axis.  All optical elements are aligned such that they are in planes perpendicular to the optical axis.

Although the measurement axis of the polarimeter is normal to its optics, the choice of the reference plane (shaded in Fig. \ref{fig:genscheme}) is arbitrary. We typically choose this plane to be aligned with the transmission axis of a calibration polarizer, through which we can pass light before it enters the polarimeter. With respect to this reference plane, the fast axis of the waveplate has an angle $\beta= \tilde \beta + \beta_0$, where $\tilde \beta$ is the angle of the fast axis of the quarter-waveplate with respect to an initially unknown offset angle $\beta_0$.
Analogously, the transmission axis of the polarizer is $\alpha=\tilde \alpha + \alpha_0$, where  $\tilde \alpha$ is the angle of the polarizer transmission axis with respect to an initially unknown offset angle $\alpha_0$. The linear polarizer transmission angle $\alpha$ is left fixed during a determination of the four Stokes parameters for the incident light, and we typically set $\tilde \alpha = 0$ so that $\alpha= \alpha_0$.  For the internal calibration procedure the linear polarizer is rotated by an angle $\tilde \alpha = 45^\circ$.  This calibration procedure (see Section \ref{sec:calib} for more details) determines $\beta_0$, $\alpha_0$ and the phase delay between the components of the light aligned with the slow and fast axes of the waveplate, $\delta \approx \pi/2$.  

Simple linear optical elements like this can be described by a Jones matrix that relates the electric field incident on the optical elements to the electric field that leaves the elements \cite{Goldstein}.  Equivalently, a Mueller matrix $\hat{M}$ transforms an input Stokes vector into the Stokes vector for the light leaving the optical elements \cite{,Goldstein},
\begin{equation}
\vec{S}_\text{out} = \hat M \,\,  \vec{S}_\text{in}.
\end{equation}
A succession of two such matrices describes the rotating waveplate polarimeter of Fig. \ref{fig:genscheme}:
\begin{equation}
\vec{S}_\text{out}=  \hat{P}(\alpha) \,\,   \hat{\Gamma}(\beta)  \,\, \vec{S}_\text{in}. 
\label{eq:soutMueller}
\end{equation}

The Mueller matrix for a waveplate whose fast axis is oriented at an angle $\beta$ with respect to a reference plane, and whose orthogonal slow axis delays the light transmission by an angle $\delta$ is \cite{Goldstein, ClarkePolLight, KligerPolLight, HandbookOptics, Shurcliff} 
\begin{widetext}
\begin{equation}
	\hat{\Gamma}(\beta) = \begin{pmatrix}  1 & 0 & 0 & 0 \\ 0 & \cos^2 2 \beta + \cos \delta \sin^2 2 \beta & \cos 2\beta \sin 2\beta (1- \cos \delta) & - \sin 2 \beta \sin \delta \\ 0 & \cos 2 \beta \sin 2 \beta (1- \cos \delta) & \cos{\delta} \cos^2 2 \beta + \sin^2 2 \beta & \cos 2 \beta \sin{\delta} \\ 0 & 
		\sin 2 \beta \sin \delta & - \cos 2 \beta \sin{\delta} & \cos{\delta}    \end{pmatrix}.
\end{equation}
The Mueller matrix for a linear polarizer \cite{Goldstein, ClarkePolLight, KligerPolLight, HandbookOptics, Shurcliff} with transmission axis oriented at an angle $\alpha$ with respect to the reference plane is 
\begin{equation}
\hat{P}(\alpha)  = \frac{1}{2} \begin{pmatrix} 1 & \cos 2 \alpha & \sin 2 \alpha & 0 \\ \cos 2 \alpha & \cos^2 2 \alpha & \cos 2 \alpha \sin 2 \alpha & 0 \\ \sin 2 \alpha & \cos 2 \alpha \sin 2 \alpha & \sin^2 2 \alpha & 0 \\ 0 & 0 & 0 & 0   \end{pmatrix}.
\end{equation}
In order to extract the Stokes parameters of the incoming light, $\vec{S}_\text{in}$, we use a photodetector to measure the intensity of the output light, $I_\text{out}$, which is up to a constant factor given by

\begin{align}
\label{eq:iout}
  I_\text{out}   (\tilde \beta)   = & I  + S \sin \delta \sin(2 \alpha_0 +2 \tilde \alpha- 2\beta_0 - 2 \tilde \beta)  \nonumber  \\   & + \,C \left[  \cos \delta \sin(2 \alpha_0 +2 \tilde \alpha- 2\beta_0 - 2 \tilde \beta)     \cos ( 2\beta_0+2  \tilde \beta)   + \cos(2\alpha_0 +2 \tilde \alpha- 2\beta_0 -2  \tilde \beta)  \sin  ( 2\beta_0+2  \tilde \beta)  \right] \\  
\nonumber   & + \,M \left[- \cos \delta \sin(2 \alpha_0 +2 \tilde \alpha - 2\beta_0 - 2 \tilde \beta)    \sin (2\beta_0+2  \tilde \beta)   + \cos(2\alpha_0 +2 \tilde \alpha- 2\beta_0-2  \tilde \beta) \cos ( 2\beta_0+2  \tilde \beta) \right].
\end{align}
\end{widetext}
The $I$, $M$, $C$ and $S$, upon which this measured intensity depends, are the Stokes parameters of the incident beam which we wish to determine. This result is in agreement with Stokes \cite{stokes1852composition} for $\tilde \beta = \beta_0 = 0$ and with  \cite{Berry:77, Romerein:11}. 

\section{Laboratory Realization and Intensity Fluctuations}
\label{sec:realization}

\begin{figure}[tb!] %[htbp!]
	\subfigure[]{\includegraphics[width=\hsize,keepaspectratio]{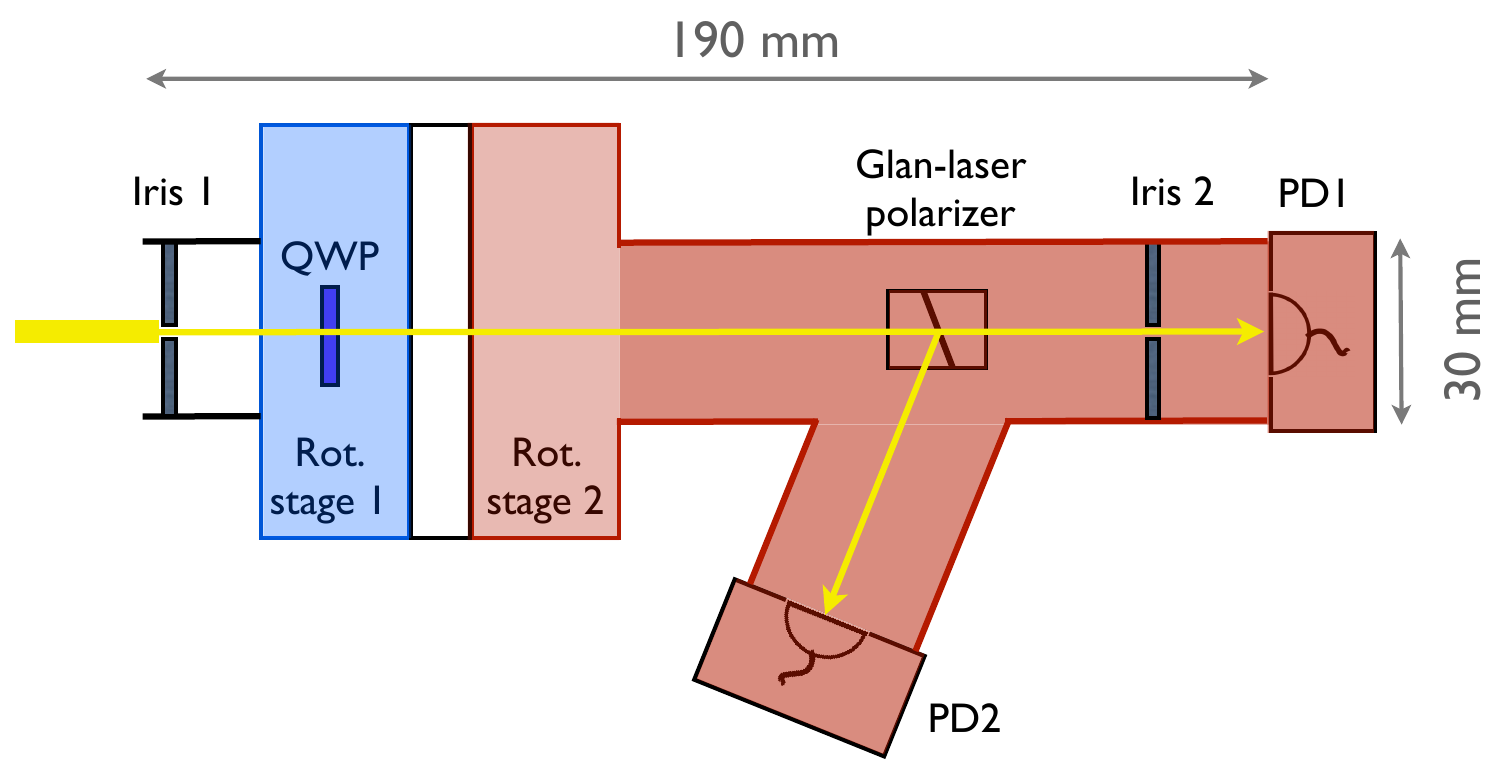}}
	\subfigure[]{\includegraphics[width=0.8\hsize,keepaspectratio]{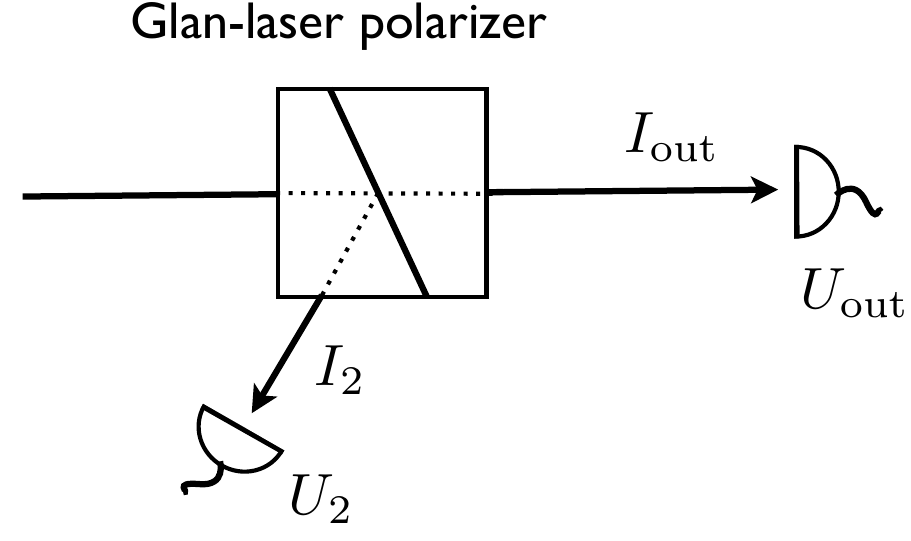}}
	\caption{(a) Scale representation of polarimeter with 1 mm apertures, a waveplate on a rotating stage, and a linear polarizer and two detectors that rotate on a second stage. (b) A Glan-laser polarizer splits the analyzed light into transmitted and refracted beams which can be used to monitor the total intensity and correct for amplitude fluctuations in the light source.}   
	\label{fig:polarScale}
\end{figure}

Our realization of a rotating waveplate polarimeter (to scale in Fig. \ref{fig:polarScale} (a)) utilizes a quarter-waveplate (Thorlabs WPQ05M-1064), a polarizer (Thorlabs GL10-B) and two detectors (Thorlabs PDA100A).  The waveplate is mounted on a first rotation stage, while the linear polarizer and the two detectors are mounted on a second rotation stage (both Newport URS50BCC).

The aperture before the polarimeter constrains the collimation of the beam inside the device. If the aperture is too large, the imperfections of optical elements (e.g. spatial inhomogeneity of the waveplate retardance) reduce the accuracy of the measurement. An aperture that is too small results in errors due to diffraction. We found that an aperture with a diameter of $1\pm0.25$ mm minimized the uncertainties for our measurements at the wavelength of 1090 nm. The second aperture is used to align the polarimeter by maximizing the light admitted by the pair of apertures.  

For measuring the polarization we typically vary the angle $\tilde \beta$ over time with 120 discretized values $\tilde \beta = \{0^\circ, 3^\circ, \dots , 357^\circ \}$  covering one full rotation of the waveplate.  We are able to similarly rotate the linear polarizer angle for the internal calibration, though a more restricted calibration rotation turns out to be optimal for reducing the uncertainties (see Section \ref{sec:calib}).

Fluctuations in light intensity contribute noise in the measured polarization since it is deduced from the intensity of light transmitted through the polarimeter.  For the sample measurements to be discussed, intensity fluctuations on the scale of few percent over the time of one polarimetry measurement contributed to fluctuations in measurements of $S/I$ of up to $\sim 2 \%$. 

To cope with these fluctuations, we implemented a normalization scheme that works at the high powers needed for our measurements.  The Glan-laser polarizer in the polarimeter (Fig. \ref{fig:polarScale} (b)) was chosen because it is suitable for our higher power requirements (unlike a Wollaston prism which has optical contacting adhesives and a lower damage threshold). This polarizer transmits light with one polarization, and sends the remaining light out through a side port where we detect it for normalization purposes.

The intensity of the incoming light source is proportional to the sum of both detector voltages. Relative gain and offset factors are applied to account for detector differences. The weighted sum signal is then used to correct the intensity of the light transmitted through the polarimeter to reduce the effect of intensity fluctuations. The calibrated offset and gain constants minimize the variation in the inferred intensity of light incident on the Glan-laser polarizer.

\section{Extracting the Stokes Parameters}
\label{sec:extractingStokes}

\begin{figure}[b]
	\includegraphics[width=0.9\hsize,keepaspectratio]{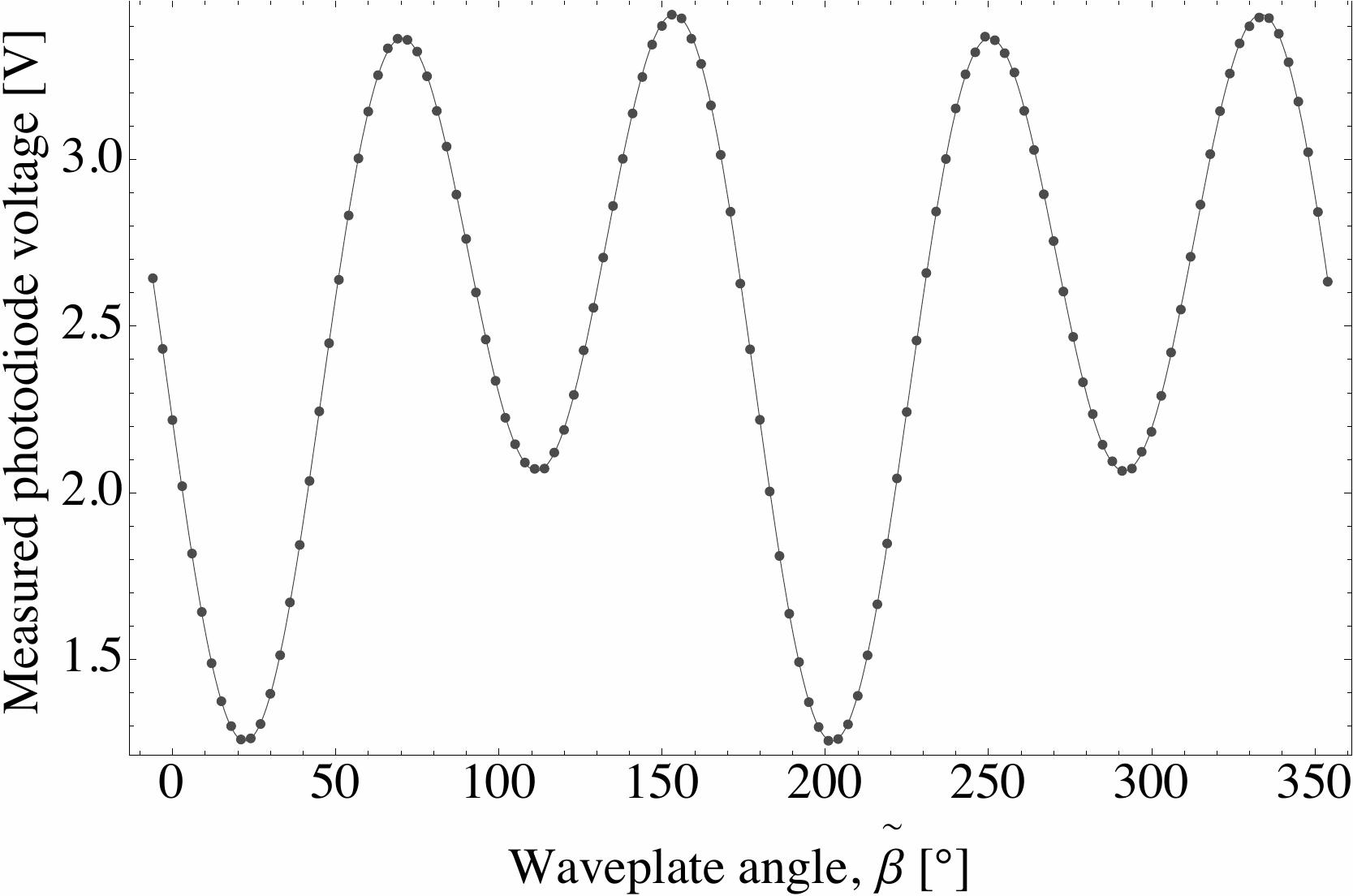}
	\caption{Illustration of how the light transmitted through the polarimeter varies with the angle of the waveplate axis as predicted.}  
	\label{fig:exampleScan}
\end{figure}

A measured signal on the polarimeter detector in Fig.~\ref{fig:exampleScan} illustrates the variation of the transmitted intensity with the waveplate angle $\tilde{\beta}$ that is described in Eq.~\ref{eq:iout}.  In terms of its Fourier components, Eq. \ref{eq:iout} can be written as

\begin{align}
\label{eq:ioutFour}
I_\text{out}  (\tilde \beta) = C_0 +C_2 \cos(2 \tilde \beta)+S_2 \sin(2 \tilde \beta) \nonumber \\
 +C_4 \cos(4 \tilde \beta)+S_4 \sin(4 \tilde \beta),
\end{align}
with the Fourier coefficients

\begin{widetext}
\begin{subequations}
\label{eq:stokesFourierCoeff}
\begin{align}
%C_0 &=  I + \frac{1+\cos (\delta)}{1-\cos (\delta)} \cdot [C_4 \cos(4\alpha_0+4 \tilde \alpha-4 \beta_0)  + S_4 \sin(4\alpha_0+4 \tilde \alpha - 4 \beta_0) ,  \\
C_0 &=  I + \frac{1+\cos (\delta)}{2} \, [M \cos(2\alpha_0+2 \tilde \alpha)  + C \sin(2\alpha_0+2 \tilde \alpha ) ],  \\
C_2 &=  S \sin \delta \sin (2\alpha_0+2 \tilde \alpha-2 \beta_0) ,\\
S_2 &= - S \sin \delta \cos (2\alpha_0+2 \tilde \alpha-2 \beta_0) ,\\
C_4 &= \frac{1-\cos (\delta)}{2} [M \cos(2\alpha_0+2 \tilde \alpha-4 \beta_0) - C \sin(2\alpha_0 +2 \tilde \alpha- 4 \beta_0)]  , \\
S_4 &=  \frac{1-\cos (\delta)}{2} [M \sin(2\alpha_0+2 \tilde \alpha-4 \beta_0) + C \cos(2\alpha_0+2 \tilde \alpha - 4 \beta_0)] .
\end{align}
\end{subequations}
These Fourier components can be extracted from measurements like the one in Fig. \ref{fig:exampleScan}. The circular polarization $S$ is related to  $S_2$ and $C_2$.  The linear polarization intensities, $M$ and $C$, are related to $C_0$, $S_4$ and $C_4$.  Inverting Eqs. \ref{eq:stokesFourierCoeff} determines the Stokes parameters of the incoming light in terms of the Fourier coefficients:
\begin{subequations}
\label{eq:stokesFourier}
\begin{align}
&I = C_0 - \frac{1+\cos (\delta)}{1-\cos (\delta)} \cdot [C_4 \cos(4\alpha_0+4 \tilde \alpha-4 \beta_0)  + S_4 \sin(4\alpha_0+4 \tilde \alpha - 4 \beta_0) ,\\
&M =  \frac{2}{1-\cos (\delta)} [C_4 \cos(2\alpha_0+2 \tilde \alpha-4 \beta_0) + S_4 \sin(2\alpha_0+2 \tilde \alpha - 4 \beta_0)] ,\\
&C =  \frac{2}{1-\cos (\delta)} [S_4 \cos(2\alpha_0+2 \tilde \alpha-4 \beta_0) - C_4 \sin(2\alpha_0 +2 \tilde \alpha- 4 \beta_0)], \\
&S =  \frac{C_2}{\sin (\delta) \sin (2 \alpha_0+2 \tilde \alpha - 2\beta_0)} = \frac{-S_2}{\sin (\delta) \cos (2 \alpha_0 +2 \tilde \alpha- 2\beta_0)} .
\end{align}
\end{subequations}
\end{widetext}
 (Note that the same Eq.\,(16) in \cite{Berry:77} has a typo in the expression for $S$.) The Stokes parameters are thus determined by the Fourier coefficients that are extracted from measurements like the one in Fig. \ref{fig:exampleScan}, along with the values of the angles $\alpha_0$, $\beta_0$ and $\delta$ from the calibration to be described.  The angle $\tilde{\alpha}$ is $0$ during a polarization measurement and is stepped away from $0$ only during a calibration, as we shall see.  

Both of the two expressions for $S$ must be used cautiously given the possibility that a denominator could vanish.  Combining them gives a more robust expression that is independent of the two calibration angles, $\alpha_0$ and $\beta_0$ \cite{Berry:77}, 
\begin{eqnarray}
\label{eq:VertS}
S  = - \rm{sign}(S_2) \frac{\sqrt{C_2^2+S_2^2}}{\sin(\delta)} . %\vert S \vert = 
\end{eqnarray}
We choose to make the angle $2(\alpha_0 - \beta_0)$ small, whereupon $\vert S_2 \vert > \vert C_2 \vert$ for nonvanishing $S$ and the sign of $S$ is that of $-S_2$.  Similarly, combining equations (\ref{eq:stokesFourier}b) and (\ref{eq:stokesFourier}c) gives a more robust expression for the magnitude of the linear polarization $L=\sqrt{M^2+C^2}$, independent of $\alpha_0$ and $\beta_0$,
\begin{eqnarray}
\label{eq:VertL}
 L= \frac{\sqrt{C_4^2+S_4^2}}{\sin^{2}{\left(\frac{\delta}{2} \right)}}. 
\end{eqnarray}
Of course, both $S/I$ and $ L/I $ still depend on all of the calibration angles since $I$ does, but the use of the more robust expression can reduce the uncertainties in $S/I$ and $L/I $.  

\section{Calibration and Uncertainties}
\label{sec:calibuncert}

The in situ calibration procedure is centrally responsible for the low uncertainties realized with this polarimeter.  The basic idea is to rotate the linear polarizer in order to calibrate the angles $\alpha_0$, $\beta_0$ and the waveplate delay $\delta$ (which varies with wavelength).  This has been done before \cite{Berry:77}.  The advance here is to determine the optimal rotation needed to minimize the calibration time and uncertainties. We avoid realignments of optical elements caused by either temporarily removing or by flipping optical elements with respect to the light transmission axis \cite{Romerein:11}.  

The uncertainties arising from the calibration cause uncertainties in the measured Stokes parameters, as do the uncertainties with which the Fourier coefficients $C_0$, $C_2$, $C_4$, $S_2$, and $S_4$ are determined.  Statistical uncertainties that can be averaged down with more measurements are typically on the order of $0.01 \%$ for $S/I$ and $0.05 \%$ for $L/I$.  %Brendon: I don't quite understand what you are saying in this sentence -- do you mean that if you average for a while you can reasonably achieve these statistical uncertainties in these quantities? Also, you haven't quite specified how you extract statistical uncertainties here.
We also discuss the systematic uncertainties that arise in addition to calibration uncertainties, specifically discussing those that arise from waveplate imperfections, misalignment of the incident light pointing relative to the measurement axis, and the finite extinction ratio of the polarizer.

\subsection{Calibration Method}
\label{sec:calib}

The calibration procedure starts with a high extinction ratio polarizer placed in the light beam before it enters the polarimeter.  The light analyzed by the polarimeter is in this case known to be almost fully polarized.  The transmission axis of this external calibration polarizer then defines the reference plane in terms of which the linear polarizations $M$ and $C$ are determined.  For the relative Stokes vector (1,0,0), Eqs. (\ref{eq:stokesFourier}) simplify to
\begin{widetext}
\begin{subequations}
\label{eq:calibD0}
\begin{align}
C_0 - \frac{1+\cos (\delta)}{1-\cos (\delta)} [C_4 \cos(4\alpha_0-4 \beta_0) + S_4 \sin(4\alpha_0 - 4 \beta_0)] & = 
 \frac{2}{1-\cos (\delta)} [C_4 \cos(2\alpha_0-4 \beta_0) + S_4 \sin(2\alpha_0 - 4 \beta_0)], \\ \arctan{\frac{S_4}{C_4}}&=2 \alpha_0 - 4 \beta_0.
\end{align}
\end{subequations}
For the third linearly independent equation needed to determine the three calibration parameters we rotate the linear polarizer from $\tilde \alpha = 0$ to  $\tilde \alpha = 45^\circ$, a choice shortly to be justified, whereupon 
\begin{align}
\tilde C_0 - & \frac{1+\cos (\delta)}{1-\cos (\delta)} \cdot [\tilde C_4 \cos(4\alpha_0+4 \tilde \alpha - 4 \beta_0) + \tilde S_4 \sin(4\alpha_0+4 \tilde \alpha - 4 \beta_0)]  \nonumber \\ = &  \frac{2}{1-\cos (\delta)} [\tilde C_4 \cos(2\alpha_0+2\tilde \alpha-4 \beta_0)  + \tilde S_4 \sin(2\alpha_0+2\tilde \alpha - 4 \beta_0)] %\\
%& \arctan{\frac{\tilde S_4}{\tilde C_4}}=2 \alpha_0  + 2\tilde \alpha- 4 \beta_0.
\label{eq:calibD45}
\end{align}
\end{widetext}
%Brendon: why do these quantities have tildes on them? It wasnt clear to me until I read the list of "steps" below.

These cumbersome equations simplify when the waveplate is reasonably close to being a quarter-waveplate, whereupon $\cos{\delta}\simeq \frac{\pi}{2}-\delta$.  They simplify further when we deliberately choose to make $\alpha_0$ small, which means that the zero-axis of the internal polarizer is approximately aligned with the external polarizer that defines the reference plane for the Stokes parameters.  The greatly simplified calibration equations are then
\begin{align}
&\delta=\frac{\pi}{2}+1+ 2 \, \frac{\sqrt{C_4^2+S_4^2}-C_0}{\sqrt{C_4^2+S_4^2}+C_0}, \\
\label{eq:caliblinAlpha0}
&\alpha_0 = \frac{\cot {\tilde \alpha}}{2} - \frac{1}{\sin (2 {\tilde \alpha} ) } \frac{ \sqrt{S_4^2+C_4^2}-\tilde C_0}{ \sqrt{S_4^2+C_4^2} -C_0 } , \\
&\beta_{0}= \frac{1}{4} \left( 2 \alpha_0 -\arctan \frac{S_4}{C_4}  \right).
\end{align}
The uncertainty in $\alpha_0$ is determined by the standard error propagation:
\begin{eqnarray}
\label{eq:uncertAlpha}
\sigma_{\alpha_0}^2 = \left( \frac{ \partial \alpha_0}{\partial {\tilde \alpha} }\right)^2 \sigma_{\tilde \alpha}^2 +\left( \frac{ \partial \alpha_0}{\partial \tilde C_0 }\right)^2 \sigma_{\tilde C_0}^2 \nonumber \\ + \left( \frac{ \partial \alpha_0}{\partial C_0 }\right)^2 \sigma_{C_0}^2 + \left( \frac{ \partial \alpha_0}{\partial C_4 }\right)^2 \sigma_{C_4}^2 + \left( \frac{ \partial \alpha_0}{\partial S_4 }\right)^2 \sigma_{S_4}^2,
\end{eqnarray}
with  $\alpha_0$ from Eq. \ref{eq:caliblinAlpha0} given our simplifying choice to make $\alpha_0$ small.

\begin{figure}[b] 
\includegraphics[width=0.9\hsize,keepaspectratio]{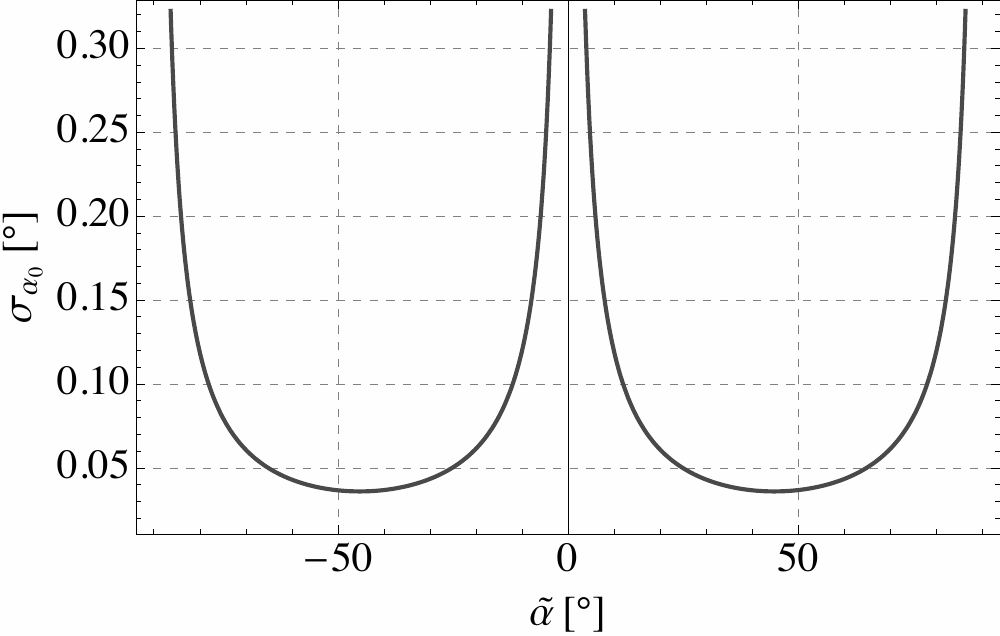}
\caption{The uncertainty with which $\alpha_0$ is determined as a function of the change in linear polarizer angle used in the calibration procedure shows that a $45^\circ$ step is close to optimal.} 
\label{fig:calib_choice}
\end{figure}

To illustrate that  the choice $\tilde{\alpha}\approx 45^\circ$ minimizes uncertainties, Fig. \ref{fig:calib_choice} shows the uncertainty in $\alpha_0$ for a typical choice of  
uncertainties in the normalized Fourier coefficients of $0.05\%$ and an uncertainty in ${\tilde \alpha}$ of $0.02^\circ$. For ${\tilde \alpha}$ close to $0^\circ$ or $90^\circ$ the equations (\ref{eq:calibD0}) and (\ref{eq:calibD45}) become degenerate and therefore the uncertainty in $\alpha_0$ grows to infinity.  The $45^\circ$ choice minimizes the uncertainties in calibration parameters, typically making them less than $0.1^\circ$, when the uncertainties in the Fourier coefficients are approximately equal.  

To summarize, these are the calibration steps:  
\begin{enumerate}
\item Place a linear polarizer before the polarimeter. Its polarization axis then defines the reference plane for $M$ and $C$. Set $\tilde \alpha=0$.
\item  Perform at least one full rotation of the waveplate recording the output intensity $I_\text{out}(\tilde \beta)$.
\item Determine the Fourier coefficients $C_0$, $C_4$, and $S_4$ from the scan in step 2.
\item Rotate the polarizer inside the polarimeter by $\tilde \alpha=45 ^\circ$ and repeat step 2. 
\item Determine the Fourier coefficients $\tilde C_0 $, $\tilde C_4$, and $\tilde S_4$ from the scan in step 4.
\item Using the Eqs. (\ref{eq:calibD0})-(\ref{eq:calibD45}), calculate the calibration parameters $\delta$, $\alpha_0$, and $\beta_0$ from the measured Fourier coefficients. %Eqs. (\ref{eq:caliblin})
\end{enumerate}

\subsection{Calibration Uncertainties}

\begin{figure*}[t]
\includegraphics[width=0.861\hsize,keepaspectratio]{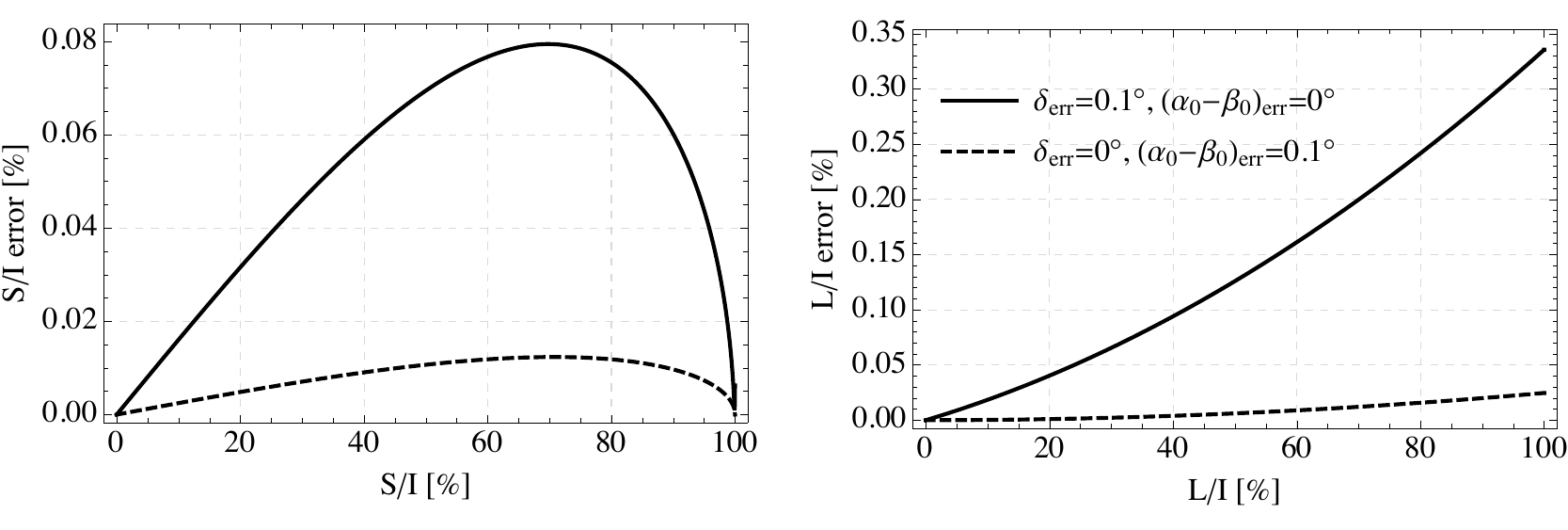}
\caption{Calculated uncertainties in $S/I$ (left) and $L/I$ (right) for typical calibration parameters of $\delta=92^\circ$, $\alpha_0=2^\circ$, $\beta_0=20^\circ$, and calibration uncertainties of $0.1^\circ$ in $\delta$ (solid curve) and also for an uncertainty of $0.1^\circ$ in $\alpha_0-\beta_0$ (dashed curve) for a range of $S/I$ and $L/I$ values. The curves for $S/I$ are symmetric around zero.}
\label{fig:calib_uncert}
\addvspace{1pt} 
\end{figure*}

Typically our calibration procedure determines the waveplate delay to about $0.1^\circ$ out of $\delta \approx 90^\circ$. For our choice of angles, $\alpha_0=2^\circ$ and $\beta_0=20^\circ$, we typically determine the differences $\alpha_0-\beta_0$ and $\alpha_0-2\beta_0$ to better than $0.1^\circ$. 

It is straightforward but tedious to propagate uncertainties of this size to resulting uncertainties in the relative Stokes parameters.  Fig. \ref{fig:calib_uncert} shows the contribution of the uncertainty in $\alpha_0 - \beta_0$ alone to errors in $S/I$ and $L/I$ with a dashed curve.  This is typically much smaller than the contribution from the uncertainty in $\delta$ alone, shown with a solid curve.  

The calibration uncertainties for $S/I$ are always below 0.1\%, and the calibration uncertainties for $L/I$ are below 0.35\% for any analyzed input polarization. For our example measurement of small $S/I$ values the calibration uncertainty is below $0.05 \%$.

\subsection{Waveplate Imperfections}

\begin{figure}[h]
\includegraphics[width=0.9\hsize,keepaspectratio]{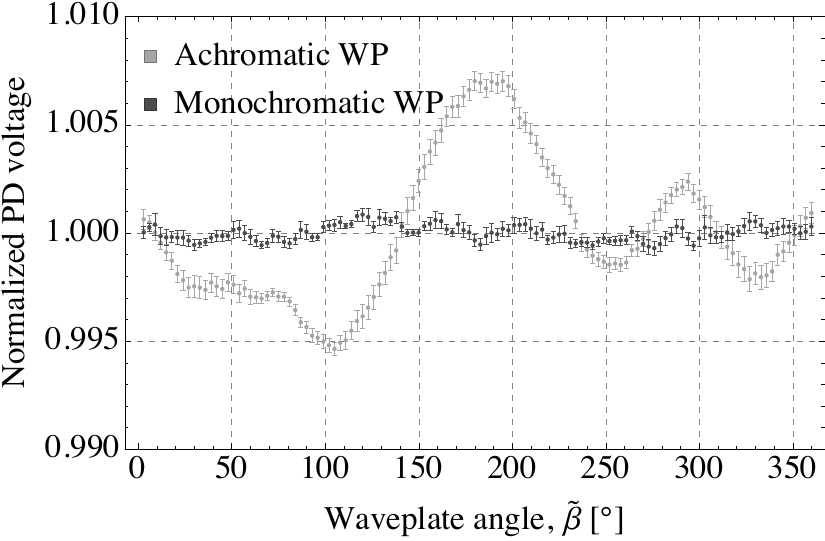}
\caption{An achromatic waveplate makes the detected intensity vary much more as a function of waveplate orientation than does a monochromatic waveplate.} \label{fig:imperfectwaveplate}
\end{figure}

Even after input intensity fluctuations were normalized out, there was still a variation in the transmitted light intensity as the waveplate rotated.  The initially observed variation, for an achromatic waveplate (Thorlabs AQWP05M-980), is shown by the light gray points in Fig.~\ref{fig:imperfectwaveplate}.  This variation limited the uncertainty in $S/I$ to about $0.3\%$.  Using instead a monochromatic waveplate (Thorlabs WPQ05M-1064) suppressed the systematic error, as shown by the dark gray points in Fig.  \ref{fig:imperfectwaveplate}.  The remaining variations typically contribute an uncertainty of less than $0.01\%$ in the normalized Fourier coefficients. This translates into an error in $S/I$ of smaller than $0.01 \%$.  

A similar systematic error was observed in astrophysical applications of rotating waveplate polarimeters \cite{Harries96, Donati98}. There, by looking at the wavelength dependence of the systematic, it was shown that the ripple of the rotating waveplate transmittance is caused by a Fabry-Perot-type interference effect. Later investigations confirmed this phenomenon \cite{Clarke05, Clarke04}.

\subsection{Misalignments}

\begin{figure}[t]
\includegraphics[width=0.95\hsize,keepaspectratio]{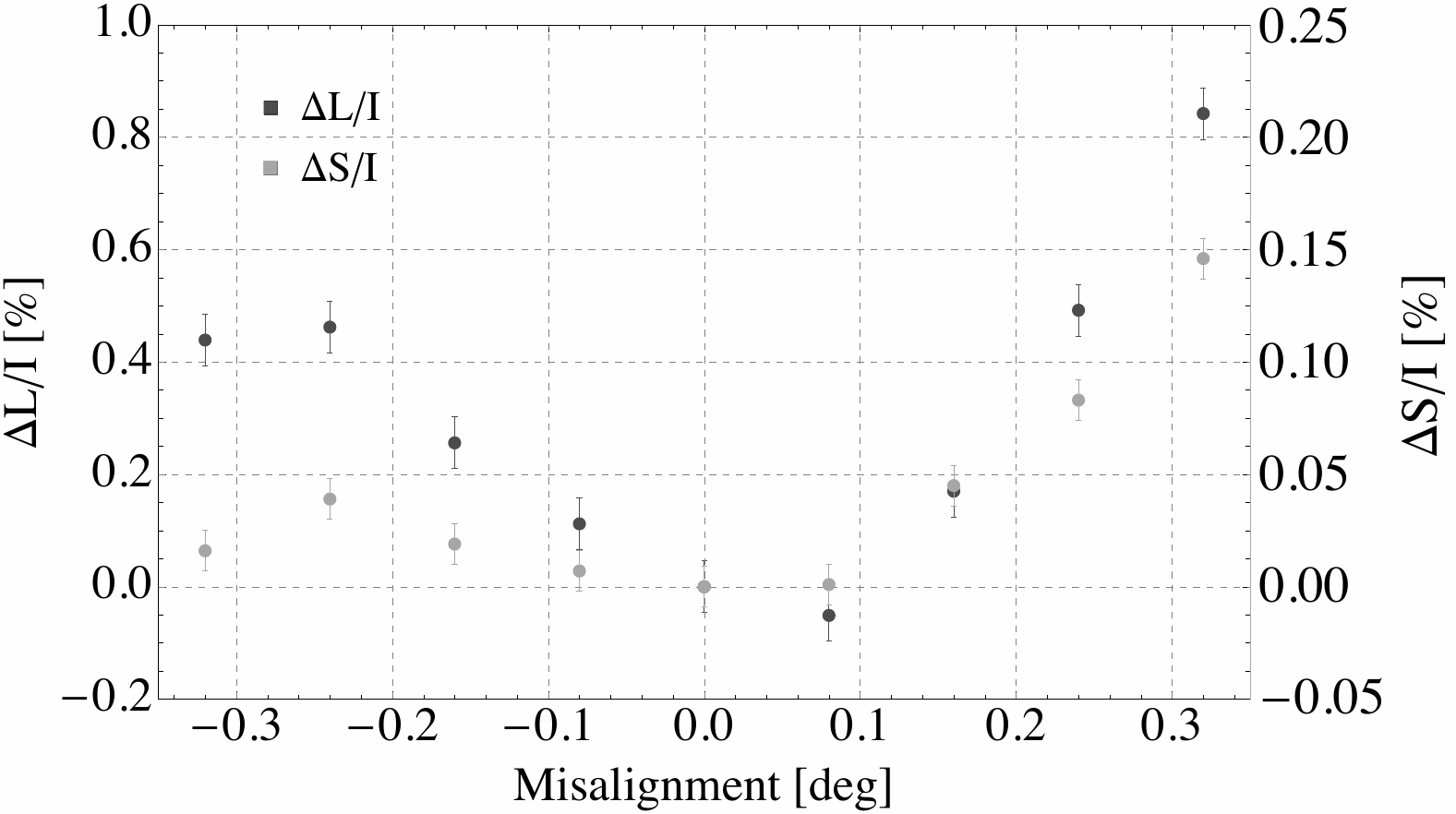}
\caption{Uncertainties in $S/I$ (light gray) and $L/I$ (dark gray) change with misalignment of the polarimeter with respect to the light propagation axis, performed with a fixed incoming polarization of $S/I =16.3 \%$ and $L/I = 98.4 \% $. }
\label{fig:misalignment}
\addvspace{-5pt} 
\end{figure}

The waveplate and the polarizer that make up the polarimeter are ideally aligned so that their optical surfaces are exactly perpendicular to the direction of propagation of the laser beam. Fig. \ref{fig:misalignment} shows an example of the systematic uncertainty that arises in measurements of $S/I$ due to misalignments.  We routinely align the polarimeter to better than $0.05^\circ$ which translates into uncertainties of $0.005 \%$ in $S/I$ and $0.05 \%$ in $L/I$.

\subsection{Finite extinction ratios of the polarizers}

The two linear polarizers used within and ahead of the polarimeter each have a finite extinction ratio $r$. A simple model of a polarizer perfectly transmits light along one axis and suppresses light transmission by a factor of $r$ along the orthogonal axis. The corresponding Mueller matrix  \cite{HandbookOptics} for such a polarizer is
\begin{equation}
\label{eq:MuellerImpPol}
\hat{P}_\text{imp}  = \frac{1}{2} \begin{pmatrix} 1 & 1-2r & 0 & 0 \\ 1-2r & 1 & 0 & 0 \\ 0 & 0 & 2 \sqrt{r (1-r)} & 0 \\ 0 & 0 & 0 & 2 \sqrt{r (1-r)}   \end{pmatrix}.
\end{equation}
For polarizers used here, $r \lesssim 10^{-5}$ rather than being perfectly $r=0$.

For calibration, the relative Stokes vector sent into the polarimeter after circular polarized light passed through an imperfect calibration polarizer is
\begin{equation}
\vec{s}= \begin{pmatrix}  1 - 2r\\ 0 \\ 2 \sqrt{r (1-r)} \end{pmatrix} \simeq  \begin{pmatrix}  1 - 2r\\ 0 \\ 2 \sqrt{r} \end{pmatrix},
\end{equation}
For an extinction ratio of $r \sim 10^{-5}$, there is a residual $S/I$ of up to $2\sqrt{r} \simeq 0.6 \%$. Numerically solving Eqs. (\ref{eq:stokesFourier}) with the residual Stokes parameters, the error on the calibration parameters is found to be less than $0.1^\circ$ for $\alpha$ and $\beta_0$, and smaller than $0.001^\circ$ for $\delta$. 

The finite extinction ratio of the polarizer in the polarimeter modifies the Fourier components measured in the out-going intensity $I_\text{out}$.  Eq. \ref{eq:iout} is re-obtained with the transformation of the Stokes parameters $M \rightarrow M(1-2r)$, $C \rightarrow C(1-2r)$ and $S\rightarrow S(1-2r)$. That means that the measured Stokes parameters differ from the true values by a factor $1/(1-2r) \simeq 1+2r$, which is $0.002 \%$ for $r \simeq 10^{-5}$.  

\subsection{Systematic Uncertainty Summary}

A summary of the investigated systematic errors for $S/I < 30\%$ and $L/I > 95 \% $, as it is in our applications, is in Table I. The leading error is due to the calibration of the retardance of the waveplate. The net uncertainty in $S/I$ is smaller than $0.1\%$. The systematic errors for $L/I$ are significantly larger, up to $0.4\%$, mainly because of higher sensitivity to the waveplate delay, $\delta$.  From Eqs. \ref{eq:VertS} and \ref{eq:VertL}, $S\propto \sin^{-1} (\delta) $ and  $ L  \propto \sin^{-2} (\delta/2) $; a small deviation from $\delta = 90^\circ$ is a first order effect in $  L  $ and is second order for  $ S  $. 

\begin{table}
%\label{table:summary}
\caption{Summary of the systematic errors for $S/I<30\%$ and $L/I > 95 \%$.}
  \begin{tabular}{ | m{4.5cm} | c  | c |}
   \hline
  Error source  & $(L/I)_{\text{err}}$ [\%] & $(S/I)_{\text{err}}$ [\%] \\
    \hline
     ($\alpha_0 -\beta_0$) calibration to $\pm 0.1 ^\circ$& $<  0.03$ &  $< 0.005$ \\ 
    $\delta$ calibration to $\pm 0.1 ^\circ$& $< 0.35$  &  $<0.05$  \\  \hline
    Intensity normalization & $  < 0.1 $ & $  < 0.02  $ \\
     Alignment of polarimeter & $  <0.05 $ & $  <0.005  $ \\
    Imperfections of waveplate & $ < 0.012  $ &$ < 0.006$ \\
    Finite extinction ratio  & $  <0.002 $ &$ <0.002 $   \\ 
  \hline  \hline
    Quadrature sum  & $ < 0.4$   & $ < 0.06$ \\
    \hline
  \end{tabular}
\end{table}

\section{Application: thermally-induced birefringence}
\label{sec:application}

To illustrate the use of our internally calibrated polarimeter we measure the circular polarization induced in laser light intense enough to create thermal gradients in glass electric field plates coated with a conducting layer of indium tin oxide used in the ACME measurement of the electric dipole moment of the electron. This effect contributed to a systematic error mechanism that dominated the systematic uncertainty in a measurement that was an order of magnitude more sensitive than previous measurements \cite{Baron:2013eja}. The polarimeter makes it possible to see whether improved electric plates produced for a second-generation experiment succeed in reducing the thermally-induced birefringence.  

To measure the polarization induced by the field plate birefringence, we start with a collimated high-power laser beam with the total power of 2~W, wavelength 1090~nm, and a circular Gaussian beam shape with waists of $w_x \simeq w_y \simeq 1.4 \, \text{mm}$. The laser beam is first polarized with the Glan-laser polarizer and then expanded in the $y$ direction using two cylindrical lenses with focal lengths of $f=10$ mm and $f=200$ mm, so that the beam shape is elongated with $w_x=1.4 \, \text{mm} \ll  w_y \simeq 30  \, \text{mm} $. The laser beam then passes through the glass plate and enters the polarimeter. $S/I$ is measured as the polarimeter is translated on a linear translation stage in the $x$ direction across the narrow illuminated area on the field plate. 

We compare the spatial gradient in $S/I$ for ACME's first- and second-generation plates.  The first-generation plate was made of borosilicate glass with a thermal expansion coefficient of $3.25 \cdot 10^{-6} \, \text{1/K}$  \cite{SchottBoro}.  The indium tin oxide layer was 200 nm thick.  The second-generation plate was designed to reduce the thermally-induced birefringence.  It is made of Corning 7980 glass with a lower thermal expansion coefficient of $0.52 \cdot 10^{-6} \, \text{1/K}$ \cite{Corning}.  To reduce absorption, the new indium tin oxide layer is thinner, at 20 nm.

\begin{figure}[t]
\includegraphics[width=0.99\hsize,keepaspectratio]{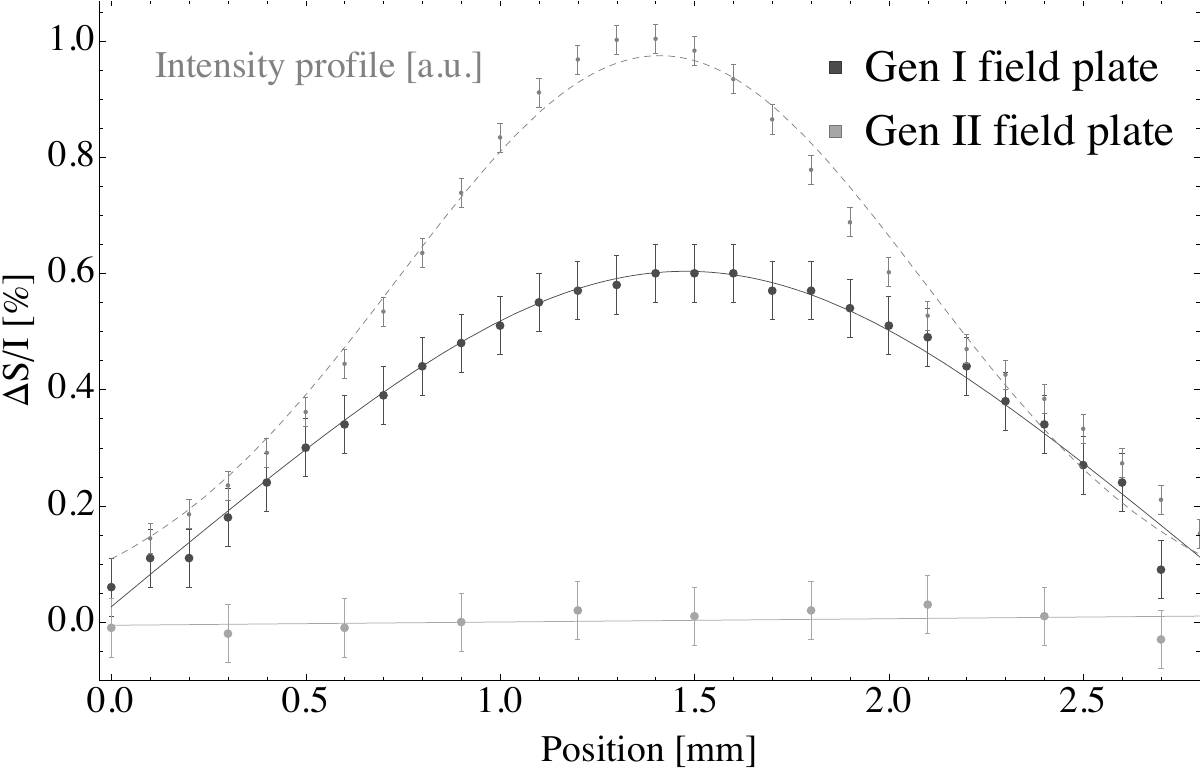}
\caption{The self-calibrated polarimeter has uncertainties low enough to compare circular polarization gradients produced by thermal gradients in first and second-generation glass field plates used by the ACME collaboration for the electron electric dipole moment experiment.  Measurements were taken with an elongated Gaussian laser beam at 1090 nm with waists $w_x=1.4 \, \text{mm} \ll  w_y \simeq 30  \, \text{mm} $ and a total power of 2 W. Error bars represent a quadrature sum of statistical and systematic uncertainties.}
\label{fig:results}
\end{figure} 

The measured changes in $S/I$ are shown in Fig. \ref{fig:results}. The intensity profile of the laser beam in the $x$ direction is the upper dashed curve.  The spatial variation of $S/I$ for the first-generation plate are the dark gray points, with a smooth curve from a theoretical model \cite{PhDNick, PhDPaul} that is beyond the scope of this report.  The much smaller spatial gradient of the light gray points was measured with the second-generation plate.  The substantial reduction, from $S/I = 0.6\%$ to $S/I < 0.1\%$, bodes very well for suppressing some systematic errors in the ACME's second-generation measurement. The small uncertainties realized with the internally calibrated polarimeter are essential for this demonstration.  

Although circular polarization gradients are less than $0.1\%$ over the diameter of the laser beam, this small variation is superimposed upon a much larger $S/I \approx 8\%$ offset.  This offset can be reduced to be less than $0.1\%$ by aligning the intensity profile of the intense laser with the polarization axis.  However, the offset is a reminder that mechanical stress in optical windows and other optical elements will typically produce birefringence.  

The $8\%$ offset in our experiment comes primarily from stress in the 5.5 x 3.5 inch vacuum windows that are 0.75 inch thick, made from the same material as the field plates.  With atmospheric pressure on both sides of these windows, adjusting the tension in screws holding the windows to the vacuum chamber changed $S/I$ from about $8 \%$ to $6 \%$.  Pumping out the chamber to put a differential pressure of one atmosphere across such a window typically changed $S/I$ by up to $3 \%$.   Related measurements with the polarimeter showed that optical elements such as a zero-order half-waveplate could produce circular polarization of up to $3 \%$. We did not observe unexpected linear polarization changes from the windows larger than the systematic uncertainties in the measurement. 

\section{Conclusion}

A highly sensitive, easy-to-construct polarimeter with high power-handling capabilities is demonstrated.  The polarimeter is calibrated internally and in situ, without the need for removing or realigning any optical elements.  The calibration procedure is critical for the low uncertainties that have been achieved.  A detailed error analysis shows that the $S/I$ Stokes parameter that describes circular polarization can be measured to better than $0.1 \%$ whereas $L/I$ is determined to below $0.4 \%$, depending on the value of $S/I$. The usefulness of a polarimeter with low uncertainty was demonstrated by measuring circular polarization gradients due to thermally-induced birefringence in a glass field plate that is critical to the most precise measurement of the electron electric dipole moment.

\section*{Acknowledgement}
This work was supported by the U.S. NSF and by the German DAAD. Our colleagues,  D. Ang, D. DeMille, J. M. Doyle, N. R. Hutzler, Z. Lasner, B. R. O'Leary,  A. D. West, E. P. West, worked with us to understand the thermally-induced birefringence in the ACME field plates and to design the improved second-generation field plates. They also made useful comments on the manuscript.

\bibliographystyle{apsrev4-1}
\bibliography{polarimeterBib}

\end{document}